\documentclass[12pt,a4paper]{article}
\usepackage{amsmath}
\usepackage{graphicx}
\usepackage{times}
\usepackage{soul}
\usepackage{cite}
\begin{document}
\begin{titlepage}
\title{The emergent black ring: a note on  increasing ratio  $\sigma_{el}(s)/\sigma_{tot}(s)$  at the LHC.}
\author{ S.M. Troshin, N.E. Tyurin\\[1ex]
\small   NRC ``Kurchatov Institute''--IHEP\\
\small   Protvino, 142281, Russia\\}

\normalsize

\date{}
\maketitle

\begin{abstract}
We discuss the  relations between the elastic and inelastic cross-sections valid for the shadow and reflective modes of the elastic  scattering. Considerations
are based on  the unitarity arguments. It is shown that the redistribution of the total interaction probability between the elastic and inelastic interactions  can lead to increasing ratio of $\sigma_{el}(s)/\sigma_{tot}(s)$ at the LHC energies in presence of the reflective scattering mode. The form of the inelastic overlap function becomes peripheral due to the negative feedback. In the absorptive scattering mode, the mechanism of this 
 increase is a different one since the impact parameter dependence of the inelastic interactions probability is  central  in this case.  A short notice is also given on the slope parameter and the leading contributions to its energy dependence in the both modes.


\end{abstract}
\end{titlepage}
\setcounter{page}{2}
\section*{Absorptive and reflective scattering modes}
It is known that the upper bounds for the inelastic cross--section \cite{mar,mar1} are different compared to the well known Froissart--Martin bound for the total cross-sections. This fact can be considered as a hint for a different asymptotic energy dependence of the inelastic cross--section compared to the total and elastic ones. Such difference indeed takes place in case of the reflective scattering mode \cite{tt,tt1,tt0} presence  at high values of the collision energy. The reflective scattering mode implies unitarity saturation at asymptotics. This mode is relevant for the energy region of very high energies.  There are strong corroborative indications of its existence
\cite{totemi, mart, csorgo} despite 
the   data analyses  at the LHC energy of $\sqrt{s}=13$ TeV \cite{totem}  are still treated as being  nonconvergent ones\cite{dremin}. 

First, we briefly mention  what the absorptive and reflective scattering modes are.
 Matrix element of the elastic scattering  is related to the corresponding partial amplitude $f_l(s)$ by the relation 
\begin{equation}\label{sm}
S_l(s)=1+2if_l(s),
\end{equation}
and the partial amplitude $f_l(s)$ is constrained by the unitarity equation:
\begin{equation}
\mbox{Im} f_l(s)=|f_l(s)|^2+h_{l,{inel}}(s)\label{ub}
\end{equation}
where
\begin{equation}
h_{l,{el}}(s)\equiv |f_l(s)|^2. \label{ubb}
\end{equation}
 Eq. (\ref{ub}) means that there are profound interrelations between elastic and inelastic interactions, see for discussion  e.g. \cite{cast}.
At high energies, it is a common practice to use an impact parameter representation\footnote{Note, that impact parameter is a conserved quantity at high energies \cite{webb}.} making replacement $l=b\sqrt{s}/2$ as well as use an assumption  on 
smallness of the
real part of elastic scattering amplitude.  As usual, we adopt this approximation for a qualitative discussion. The arguments \cite{white, dremr}  are based on the theoretical results and numerical estimations. The assumption is in agreement with the small value of the ratio of the real to imaginary part of the forward scattering amplitude \cite{totro}. And corollary to that, the unitary upper bound for $[\mbox{Re} f(s,b)]^2$ is only 1/4 of the corresponding bound for
 $[\mbox{Im} f(s,b)]^2$.

Unitarity  written in the impact parameter representation (the real part of the scattering amplitude
 is neglected) provides relations for the  differential distributions  of the elastic and inelastic collisions over $b$:
\begin{equation}\label{el}
h_{el}(s,b)= f^2(s,b),
\end{equation}
\begin{equation}\label{inel}
h_{inel}(s,b)= f(s,b)(1-f(s,b)).
\end{equation}
Thus, the  amplitude $f(s,b)$ variation is limited by the values from the interval
$0 \leq f \leq 1$ and $f=1/2$ means the complete absorption of the initial state, i.e. $S=0$, and $h_{el}\leq 1$, but $h_{inel}\leq 1/4$. 

Absorptive scattering mode corresponds to the interval  $0< f \leq 1/2$, while the reflective scattering mode appears in the region of   $s$ and impact parameters $b$ such that the amplitude $f$ acquires values in the range of $1/2 < f \leq 1$.  As it follows from the analysis \cite{mart}, one can safely assume that the elastic scattering has an absorptive nature in the energy region $\sqrt{s}\leq 5$ TeV, i.e. in this energy region: $f\leq 1/2$ at all values of the impact parameter $b$ and
the following inequality takes place since $f\leq 1/2$:
\begin{equation}\label{ineqo}
h_{el}(s,b)\leq h_{inel}(s,b)\leq 1/4,
\end{equation}
where $h_{el,inel}$ are the respective overlap functions. 
Therefore, in the above energy region, the elastic and inelastic cross--section should  obey the relation
\begin{equation}\label{ineq}
\sigma_{inel}(s)\geq \sigma_{el}(s),
\end{equation}
since 
\begin{equation}\label{ineqi}
\sigma_{el,inel}(s)=8\pi \int_0^\infty bdb h_{el,inel}(s,b).
\end{equation}
Eq. (\ref{ineq}) is  an important inequality for discussion of the absorptive versus reflective modes, but not quite new,
e.g. the Pumplin bound \cite{pump} for inelastic diffraction presupposes such a relation between elastic and inelastic cross-sections since it has been obtained in a absorptive approach\footnote{Generalized upper bound  when both -- absorptive and reflective -- modes  are presented has been obtained in \cite{jpg}.}. 

Thus, the absorption is the reason for the inelastic interactions dominance. But the opposite claim is not valid, from dominance of the inelastic interactions one cannot conclude that the absorption only  takes place in the {\it whole region of the impact parameter variation}.

When the energy becomes greater than the threshold value
$s_r$\footnote{The threshold value  ${s_r}$ is determined by $S(s_r,b=0)=0$.} , the scattering picture at small values of the impact parameter 
($b\leq r(s)$, where  $S(s,b=r(s))=0$ and $S(s,b)$ becomes negative at $ b\leq r(s)$\footnote{Note, that $r(s)\sim \ln s$ at $s\to\infty$  \cite{tt1}.}), starts to acquire a reflective contribution and the Eq. (\ref{ineq})  transforms into the following two inequalities:
\begin{equation}\label{ineqor}
\int_{r(s)}^{\infty}bdb h_{el}(s,b)<\int_{r(s)}^{\infty}bdb h_{inel}(s,b)
\end{equation}
and
\begin{equation}\label{ineqorc}
\frac{r^2(s)}{2}\geq\int_{0}^{r(s)}bdb h_{el}(s,b)>\frac{r^2(s)}{8}>\int_{0}^{r(s)}bdb h_{inel}(s,b).
\end{equation}
 It happens due to unitarity constraint for $h_{el}$ and appearance of the new relation between $h_{el}$ and $h_{inel}$ at  $ b< r(s)$, i.e.
\begin{equation}\label{ov}
1\geq h_{el}(s,b)>1/4>h_{inel}(s,b),
\end{equation}
(see Fig. 1). Eq. {\ref{ineqorc}} leads to the lower bound for integrated elastic cross--section, $$\sigma_{el}> \pi r^2(s)$$ and (since $\sigma_{tot}(s)=\sigma_{el}(s)+\sigma_{inel}(s)$) to the upper bound for the total cross--section of inelastic interactions, $$\sigma_{inel}(s)<\sigma_{tot}(s)-\pi r^2(s).$$
The above heuristic inequalities are valid at finite energies, $s>s_r$, the rigorous asymptotic ones have been obtained in \cite{mar,mar1}.
\begin{figure}[hbt]
	\vspace{-0.42cm}
	\hspace{-1cm}
	\resizebox{15cm}{!}{\includegraphics{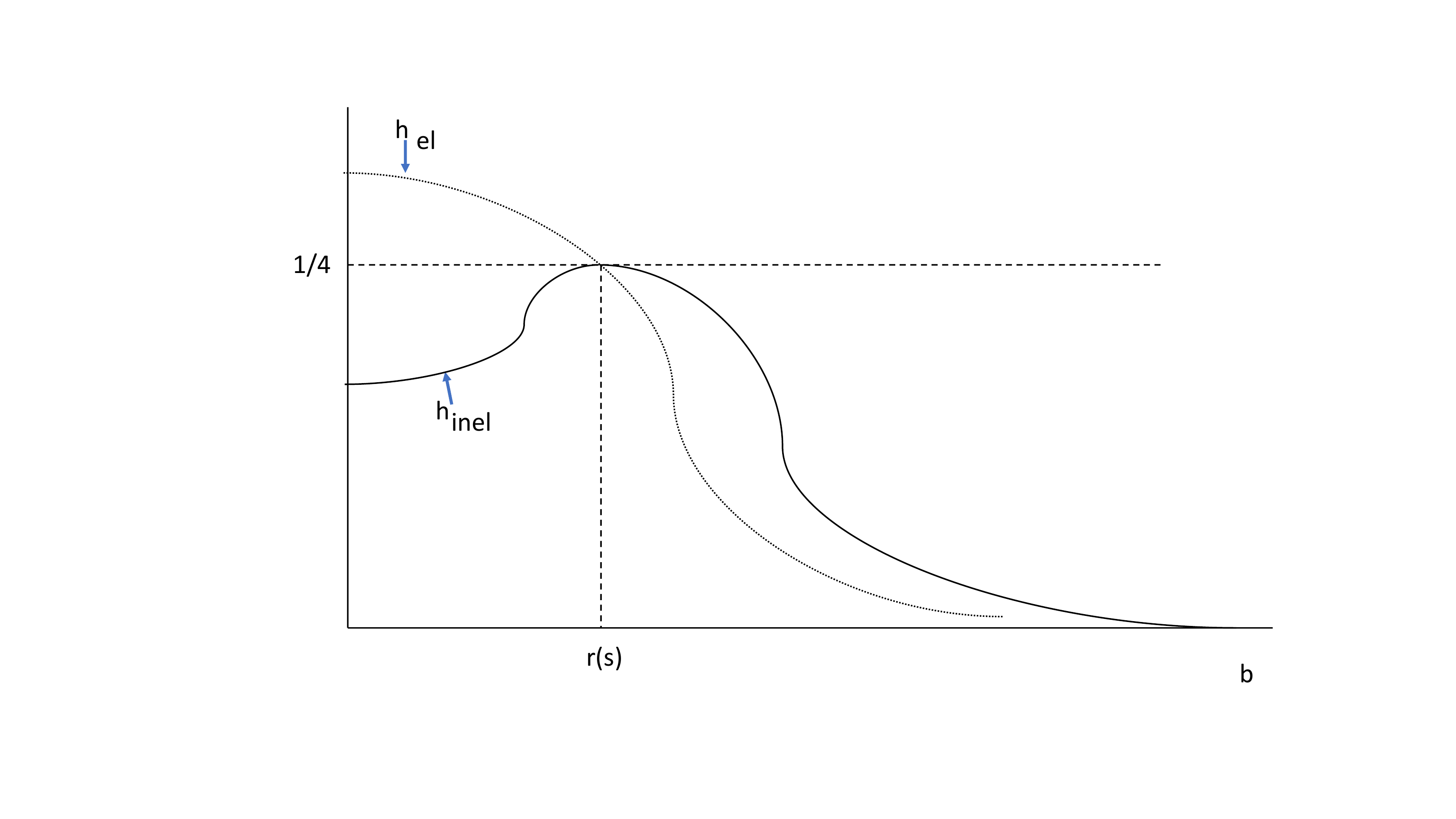}}		
	\vspace{-1cm}
	\caption{Schematic representation of the overlap functions $h_{el, inel}$ impact parameter dependencies at the energy $s>s_r$.}	
\end{figure}	 

The illustrative schematic representation of Fig. 1 reflects qualitatively the picture corresponding to the quantitative, model--independent impact parameter analysis performed in \cite{totemi,mart,csorgo}.

To  confront above inequalities  with the data, the explicit forms of the impact parameter dependencies of the functions $h_{el,inel}$ are needed. Currently, there are uncertainties  under extraction of such information  from the available data. But it should be noted here that analysis performed in \cite{totemi} gives the following values $h_{el}(s, b=0)=0.31$ and $h_{el}=1/4$ at $b \simeq 0.35$ fm at the energy $\sqrt{s}= 8$ TeV  for the central form of the impact parameter dependence of elastic overlap function (cf. Fig.19   of ref. \cite{totemi}). 
\section{Central versus peripheral behavior  of $h_{el}$ and  $h_{inel}$}
Fig.19   of ref. \cite{totemi} includes the second curve corresponding to a peripheral impact parameter dependence of $h_{el}(s,b)$ implying that elastic scattering amplitude according to unitarity (cf. Eqs.{(\ref{ub})-(\ref{inel})}) is determined by elastic overlap function in the region of large impact parameters. But, the scattering amplitude
(both the real and the imaginary parts of the amplitude) is exponentially decreasing with $b$  in the region of large impact parameters due to analyticity and becomes small. Such decreasing dependence leading to smallness of the amplitude combined with unitarity contradicts to a peripheral form of the impact parameter dependence of the elastic overlap function $h_{el}(s,b)$ since  at large values of impact parameter  the scattering amplitude is determined  by the inelastic collisions according to unitarity:
\begin{equation}
f\simeq h_{inel},
\end{equation}
while $h_{el}(s)=o(f)$ at large values of $b$.
The option of central inelastic overlap function and peripheral elastic overlap function is not considered therefore. 

Thus, we have $f(s,b=0)=0.56$ and $h_{inel}(s,b=0)=0.248$ (i.e. shallow minimum at $b=0$) and $h_{inel}(s,b=0)<h_{inel}(s,b \simeq 0.35)=1/4$ \cite{totemi}. Hence, already at the LHC energy $\sqrt{s}= 8$ TeV the black disk limit is exceeded at small values of collision impact parameter. Due to monotonic energy dependence of the scattering amplitude the effect should be  more pronounced at $\sqrt{s}= 13$ TeV. Such black disk limit exceeding has been discussed  more than 25 years ago \cite{tt}. There are, of course,  ambiguities  in extraction of $b$--dependent amplitude of elastic scattering from experimental data related to the presence of the two signs of a
 square root of differential cross--section (cf. \cite{dremr}), but no one seems to be objective to the principal statement on the reaching a black disc limit at $b=0$ at the LHC energies.  An incomplete list of the relevant references includes the following ones \cite{mart, csorgo, dremr, anis, bron}. 

We  adopt the point that this limit has already been  exceeded at the LHC energies and represent then the scattering amplitude $f(s,b)$ in the region $0 \leq b \leq r(s)$ and a  fixed the LHC  energy  in the form
\begin{equation}
\label{alp}
f(s,b)=\frac{1}{2}[1+\alpha (s,b)]
\end{equation}
with small positive function $\alpha (s,b)$ such that $\alpha (s,b=r(s))=0$. The value of $\alpha$ at $b=0$ and $\sqrt{s}=8$ TeV is of order of 10 \% \cite{totemi}. The elastic overlap function can be written at the LHC energies and $b\leq r(s)$ in the form 
\begin{equation}
\label{hel}
h_{el}(s,b)=\frac{1}{4}[1+2\alpha (s,b)+\alpha^2 (s,b)],
\end{equation}
while inelastic overlap function has a negative deviation (hollowness) of the second order on $\alpha$:
\begin{equation}
\label{hinel}
h_{inel}(s,b)=\frac{1}{4}[1-\alpha^2 (s,b)]. 
\end{equation}
Thus, it is evident that due to a smallness of deviation from the black disc limit\footnote{This limit corresponds to a complete absorption of the initial state} at the LHC energies  exceeding of this limit  should  be detected more easily  under the studies of the elastic overlap
overlap function than under  the studies of the  inelastic one at the LHC energies. Due to the black disk limit exceeding elastic overlap function gets a positive feedback and the inelastic one gets a negative feedback.

Eq. (\ref{ineqorc}) can be used for a qualitative explanation of the ratio  increase with energy at the  LHC \cite{totemr} where the reflective scattering mode seems to appear. 
Indeed, we observe the mentioned
redistribution of the total probability increase with the energy growth in favor of the elastic interactions under transition to a reflective mode (see Fig. 1, Eqs. (\ref{hel}) and (\ref{hinel})). In this regard, it is helpful to recall that 
$\sigma_{el}(s)\sim \ln^2 s $, but $\sigma_{inel}(s)\sim \ln s $ and $\sigma_{el}(s)/\sigma_{tot}(s)\to 1$ at $s\to\infty$ \cite{tt1}.

It also is  instrumental to keep in mind that in the reflective scattering mode $S<0$. The following relations are relevant for clarification of the energy dependencies of the functions  $h_{el,inel}$ \cite{tt0}:
\begin{equation} \label{elo}
\frac{\partial h_{el}(s,b)}{\partial s}=(1-S(s,b))\frac {\partial f(s,b)}{\partial s}
\end{equation}
and 
\begin{equation} \label{inelo}
\frac{\partial h_{inel}(s,b)}{\partial s}=S(s,b)\frac {\partial f(s,b)}{\partial s}.
\end{equation}
Thus, the following limit takes place, namely, $h_{inel}(s, b=0)\to 0$ when $f(s,b=0)$ approaches unitary  limit, i.e. $f(s,b=0)\to 1$ at $s\to\infty$.

It should also be noted that the above mechanism is not relevant for the absorptive scattering mode where the inelastic overlap function has a central impact parameter dependence. At the energies where only absorptive scattering takes place\footnote{The CERN ISR energies are in the absorptive scattering region \cite{amaldi}.} the ratio $\sigma_{el}(s)/\sigma_{tot}(s)$ increase is due to a faster increase of elastic scattering probability at small and moderate values of impact parameter compared to the probability of collision with the same values of the impact parameter leading to the inelastic interactions.  

\section{The slope parameter}

These results have some implications for the slope parameter $B(s)$ of the forward elastic peak,
\begin{equation}\label{bs}
B(s)=\frac {d}{dt} \ln \frac{d\sigma}{dt}|_{t=o}.
\end{equation}
This quantity is determined by the average value of the impact parameter squared, i.e.
\begin{equation}\label{ineqb}
\langle b^2 \rangle _{tot} \equiv {\int_0^\infty b^2f(s,b)bdb}/{\int_0^\infty f(s,b)bdb}.
\end{equation}
 In its turn, according to unitarity the energy dependence of $\langle b^2 \rangle _{tot}$ is determined  by the ones of cross--sections $\sigma_{el,inel}$ and average values $\langle b^2 \rangle _{el,inel}$.

In the absorptive scattering mode the major contribution to $B(s)$ comes from inelastic processes:
\begin{equation}\label{ineqn}
\sigma_{inel}(s)\langle b^2 \rangle _{inel}\geq\sigma_{el}(s)\langle b^2 \rangle _{el},
\end{equation}
where
\begin{equation}\label{ineqq}
\langle b^2 \rangle _{el,inel} \equiv {\int_0^\infty b^2h_{el,inel}(s,b)bdb}/{\int_0^\infty h_{el,inel}(s,b)bdb}.
\end{equation}
But the elastic interactions give a subdominant contribution to the slope parameter in this mode.
The relation  Eq. (\ref{ineqn}) is a direct result of Eq. (\ref{ineqo}).

Contrary, in the reflective scattering mode, due to unitarity saturation, the main contribution to $B(s)$ comes from the elastic scattering {\it decoupled from the inelastic interactions} \cite{epl} and   the following limiting dependence at $s\to\infty$ should take place due to the  elastic scattering asymptotic dominance:
\begin{equation}\label{bsa}
B(s)\to \langle b^2 \rangle _{el}/2.
\end{equation}
It should be noted again that elastic scattering dominance at $s\to\infty$ is a natural result of a self--damping of the  inelastic channels' contributions in this limit \cite{baker}, i.e. it is a consequence of the aforementioned redistribution between the probabilities of elastic and inelastic interactions due to unitarity.  At the same time, the average values  $\langle b^2 \rangle _{el,inel}(s)$ have similar dependence in the limit of $s\to\infty$ \cite{epl}:  $$\langle b^2 \rangle _{el,inel}(s)\sim r^2(s).$$

Redistribution of probabilities with corresponding decrease with energy   of the inelastic interactions' probability at small impact parameters is a result of the reflective scattering mode appearance at the LHC energies. One could expect a speed up of the ratio  $\sigma_{el}(s)/\sigma_{tot}(s)$ increase  compared to the absorptive mode at lower energies due to starting decrease with energy of the inelastic interaction probability at the point of $b=0$ and in its vicinity   ($S<0$). The region  where reflective scattering mode is expected to give a noticeable contribution corresponds to the LHC energy region. Contrary, one should expect a slow down of the ratio  $\sigma_{el}(s)/\sigma_{tot}(s)$ increase at the LHC energies in case of asymptotic saturation of the black disk limit at $s\to\infty$ (cf. \cite{fag}).

\section{Discussion}
The mode is called reflective 
since the phases of incoming and outgoing  waves differ by $\pi$. In optics, its appearance  is associated with  density increase of a reflecting medium beyond some threshold value. Such medium has  a higher refractive index than the  one  incoming wave arrived from. In QCD, one can assume that the color conducting medium is formed in the transient state of hadron interactions and it is responsible for reflective scattering mode \cite{interp}. There is a link between structure of proton responsible for reflective scattering and that of the chiral models. The chiral models describe baryon as consisting from an inner core with a baryonic charge and an outer cloud surrounding the core.  Presence of the inner repulsive core is in agreement with the DVCS data analysii \cite{burk} and with the indirect (presence of the second exponential slope in $d\sigma/dt$) LHC data at $\sqrt{s}=13$ TeV \cite{totem}. These results are in favor of the two--scale structure of proton. The interpretations of the results of the CLAS detector at Jefferson Laboratory based on the existence of the extended substructures inside the proton was discussed also in \cite{petr}.

The reflective scattering mode appears when the value of elastic overlap function exceeds the black disk limit, i.e. it becomes greater than 1/4 at $b=0$. Due to unitarity relation in the impact parameter representation the value of the inelastic overlap function becomes less than 1/4 at $b=0$. The latter phenomenon was called hollowness and  widely discussed in the recent papers. We mention only some of them here, i.e. \cite{bron, bron1, soto,soto1}. It is evident that hollowness and reflective scattering are, in fact, represent the ``two sides of the same medal'', those are related by unitarity and appear when the black disc limit is exceeded.  The value of the ratio $r_e(s)\equiv\sigma_{el}(s)/\sigma_{tot}(s)$ at $s>s_r$ is correlated with the degree of reflection (albedo in optics), while the value of the ratio $r_i(s)\equiv\sigma_{inel}(s)/\sigma_{tot}(s)$\footnote{The ratio $r_i(s)$ can be related to the real to imaginary part ratio of forward scattering amplitude \cite{impl} due to the local dispersion relations  \cite{bronz} .}  is correlated with the degree of the hollowness. The sum of these two ratios $r_e+r_i=1$ due to unitarity. Discussion of relation of the reflective scattering with a peripheral form of the inelastic overlap function (hollowness) and the particular effects of the reflective scattering mode in hadron production can be found e.g.  in \cite{trt}. The peripheral form of $h_{inel}(s,b)$ would lead, in particular, to slow down  of the mean multiplicity growth at the LHC \cite{trt1}.

The appearance of these new twin phenomena (reflective scattering and hollowness) is a manifestation of the fact that the restricting assumption on the sole absorption  is not an equivalent to effective fulfillment of unitarity \cite{tt1, baker,  sach} and the particular  scattering amplitude unitarization procedure should  reproduce the {\it  both} scattering modes --- the absorptive and the reflective ones ---- under description of the energy evolution of the scattering amplitude. Constraining   assumption on the existence of the single scattering mode 
(absorption only)  has no firm physical ground.  Reflective scattering and hollowness can be associated with an effect of the self--dumping intermediate inelastic channels contribution \cite{baker} at the LHC energies. The self-dumping can arise as a result of randomization of the phases of  multiparticle production amplitudes.  This  randomization in its turn can be considered as a consequence of the color--conducting collective state of hadronic matter formation under the central over impact parameter hadron collisions  (with high multiplicities) and  subsequent stochastic decay of this state and hadronization  into the multiparticle final states \cite{interp}.  Important role of the phases for the inelastic overlap function dependence has been discussed in \cite{koba}. Their mutual cancellation has been proposed as an explanation for the diffraction peak appearance in elastic scattering \cite{pred}.

It is also useful here to pay an attention again to the results of the seminal paper \cite{gold} on the alarming tendencies in the analytical properties of the scattering amplitude observed in  the approach based on the exponential unitarization when a simple pole singularity  in a phase function  of energy turns into an essential singularity in the scattering amplitude after the unitarization procedure.

\section*{Acnowledgements}
We are grateful to T. Cs\"{o}rg\H{o},  J. Ka\v{s}par and E. Martynov for the interesting and useful correspondence on the amplitude impact parameter analysis.


\begin{thebibliography}{99}
\bibitem{mar}
A. Martin, Phys. Rev. D {\bf 80}, 065013 (2009).
\bibitem{mar1}
T.T. Wu, A. Martin, S.M. Roy, and V. Singh,  Phys. Rev. D {\bf 84},  025012 (2011).
\bibitem{tt}
S.M. Troshin, N.E. Tyurin, Phys. Lett. B {\bf  316}, 175 (1993).
\bibitem{tt1}
S.M. Troshin, N.E. Tyurin, Int. J. Mod. Phys. A {\bf 22}, 4437 (2007).
\bibitem{tt0}
S.M. Troshin, N.E. Tyurin, Phys.  Rev. {\bf 88}, 077502 (2013). 
\bibitem{totemi}
G. Antchev et al. (TOTEM Collaboration), Eur. Phys. J., C {\bf 76}, 66 (2016).
\bibitem{mart}
A. Alkin, E. Martynov, O. Kovalenko, S.M. Troshin, arXiv: 1807.06471v2.
\bibitem{csorgo}
T. Cs\"{o}rg\H{o}, R. Pasechnik, A. Ster,
EPJ Web Conf. {\bf 206}, 06007 (2019).
\bibitem{totem}
G. Antchev et al. (TOTEM Collaboration),  CERN-EP-2018-338; arXiv:1812.08283.
\bibitem{dremin}
I.M. Dremin, V.A. Nechitailo, Eur. Phys. J.  C {\bf 78}, 913 (2018) .
\bibitem{cast}
R. Castaldi, G. Sanguinetti, Ann. Rev. Nucl. Part. Sci., {\bf 35}, 351 (1985).
\bibitem{webb}
B.R. Webber, Nucl. Phys. B {\bf 87}, 269  (1975).
\bibitem{white}
I.M. Dremin, V.A. Nechitailo, S.N. White, Eur. Phys. J. C {\bf 77 }, 910 (2017).
\bibitem{dremr}
I.M. Dremin,  Particles  {\bf 2}, 57 (2019).
\bibitem{totro}
G. Antchev et al. (TOTEM Collaboration), arXiv: 1812.04732; CERN-EP-2017-335; CERN-EP-2017-335-v3.
\bibitem{pump}
J. Pumplin, Phys. Rev. D {\bf 8}, 2899 (1973).
\bibitem{jpg}
S.M. Troshin, N.E. Tyurin, J.  Phys. G {\bf 44}, 015003 (2017).
\bibitem{anis}
V.V. Anisovich, V.A. Nikonov, J. Nyiri,
Phys.Rev. D {\bf 90} , 074005  (2014).
\bibitem{bron}
W. Broniowski et al. Phys. Rev. D {\bf 98}, 074012 (2018).
\bibitem{amaldi}
U. Amaldi, K.R. Shubert, Nucl. Phys. B {\bf 166}, 301 (1980).
\bibitem{epl}
S.M. Troshin, N.E. Tyurin, Eur.  Phys. Lett. {\bf 124},   21001 (2018).
\bibitem{totemr}
G. Antchev et al. (TOTEM Collaboration), CERN-EP-2017-321; CERN-EP-2017-321-V2; arXiv:1712.06153.
\bibitem{baker}
M. Baker, R. Blankenbecler, Phys. Rev. {\bf 128}, 415 (1962).
\bibitem{fag}
D.A. Fagundes, M.J. Menon, P.V.R.G. Silva, Nucl. Phys. A {\bf 946}, 194 (2016).
\bibitem{interp}
S.M. Troshin, N.E. Tyurin, J. Phys. G: Nucl. Part. Phys. https://doi.org/10.1088/1361-6471/ab0ed1, in press.
\bibitem{burk}
V.D. Burkert, L. Elouadrhiri, F.X. Girod, Nature {\bf 557}, 396 (2018).
\bibitem{petr}
R. Petronzio, S. Simula, G. Ricco, Phys. Rev. D {\bf 67}, 094004 (2003).
\bibitem{bron1}
E. Ruiz Arriola, W. Broniowski, Phys. Rev. D {\bf{95}}, 074030 (2017).
\bibitem{soto}
J. L. Albacete, A. Soto-Ontoso, Phys. Lett. B {\bf 770}, 149 (2017).
\bibitem{soto1}
J.L. Albacete, H. Petersen, A. Soto-Ontoso,  Phys.Rev. C {\bf 95}, 064909 (2017).
\bibitem{impl}
S.M. Troshin, N.E. Tyurin,  Mod. Phys. Lett. A {\bf 33}, 1850206 (2018).
\bibitem{bronz}
J.B. Bronzan, G.L. Kane, U.P. Sukhatme, Phys. Lett. B {\bf {49}}, 272 (1974).
\bibitem{trt}
S.M. Troshin, N.E. Tyurin, Int. J.  Mod. Phys. A {\bf 29}, 1450151 (2014).
\bibitem{trt1}
S.M. Troshin, N.E. Tyurin,  Phys. Lett.  B {\bf 732}, 95 (2014).
\bibitem{sach}
C.T. Sachrajda, R. Blankenbecler, Phys. Rev. D. {\bf 12}, 1754 (1975).
\bibitem{koba}
Z. Koba, M. Namiki, Nucl. Phys. B {\bf 8}, 413 (1968).
\bibitem{pred}
V. Barone, E. Predazzi, { High-Energy Particle Diffraction},
{\it Berlin Heidelberg: Springer-Verlag (2002)}.
\bibitem{gold}
R. Blankenbecler, M.L. Goldberger, Phys. Rev. {\bf 126}, 766 (1962).
\end{thebibliography}
\end{document}